\documentclass{pinchcr}
\usepackage{amsmath}
\usepackage{graphics}
\usepackage{graphicx}
\usepackage{amssymb}
\usepackage{setspace}
\usepackage{bm}
\usepackage{ragged2e} 
\usepackage{fancyhdr}

\fancypagestyle{plain}{
	\fancyhf{}
	\fancyhead[LE,RO]{\thepage}
	
}

\title{Temperature Dependence of the Entropy and Magnetic Moment of Landau--Quantized Dichalcogenide Carriers in a Strong Magnetic Field}
\author{Norman J. M. Horing \\
	Department of Physics \\
	Stevens Institute of Technology \\
	Hoboken, NJ 07030, USA \\
	17 January 2019 \\ \ \\
	\textbf{ABSTRACT} \\ 
	\begin{flushleft}
		\justify\normalsize
		\qquad This work addresses the temperature dependence of the statistical thermodynamic functions of the Group VI Dichalcogenides in a quantizing magnetic field. These functions include the grand partition function, the ordinary partition function, the Helmholtz Free Energy and the entropy: Their dependencies on temperature and magnetic field strength are carefully examined, particularly in the degenerate regime and the approach to zero temperature, also in the nondegenerate regime. The joint dependence on magnetic field and temperature is aso examined in the Dichalcogenide magnetic moment in both statistical regimes.
	\end{flushleft}}
\begin{document}
	\let\cleardoublepage\clearpage
	\pagenumbering{arabic}
	\maketitle
	
	\pagestyle{plain}
	
	\chapter{Introduction}
	\indent This paper addresses some fundamental physical properties of ``Dirac--like" materials, in particular the group VI Dichalcogenides. Starting with Graphene$^{1-6}$, such materials have been at the focus of research attention since the discovery of the extraordinary electrical conduction and detection properties of Graphene about fifteen years ago. Additional materials of this type include Silicene$^7$, Topological Insulators$^8$, as well as the Dichalcogenides$^9$ (and some others). All are under intense investigation worldwide in all science and engineering disciplines for their potential to succeed Silicon as the material of choice for the next generation of electronic devices and computers. Recognition of the importance of these materials has been underscored by the award of the 2010 Nobel Prize in Physics to Geim and Novoselov for their pioneering work on Graphene. The fact that the low energy carrier spectrum of``Dirac--like" materials mimics that of relativistic electrons/positrons (with energy proportional to momentum) also heightens intellectual interest in them as accessible solid state laboratories of relativistic physics, albeit with different parameters. Much has already been done in regard to experimental and theoretical studies on Graphene, and considerable work on the other ``Dirac--like" materials is now filling the scientific and engineering literature. \\
	\indent Here, we address the thermodynamic properties of Group VI Dichalcogenides$^9$ to study their entropy subject to Landau quantization in a high magnetic field. The entropy, $S$, is determined by a variation of the Helmholtz Free Energy, $F$, in the thermodynamic relation
	\begin{equation}
		dF = -pdV - SdT' + \mu dN
	\end{equation}
	($p$ is pressure (lineal in 2D), $\mu$ is chemical potential, $T'$ is Kelvin temperature, $N$ is number and $V$ is volume (area in 2D)). Holding $N$ and $V$ constant, the entropy may be identified as
	\begin{equation}
		S = -(\partial F/\partial T')_{N, V}
	\end{equation}
	with density constant. Clearly, it is important to examine the temperature dependence of the thermodynamic functions carefully for an accurate determination of their entropy. Moreover, in view of the great importance of the magnetic field as an agent for probing the properties of matter and also modifying them$^{10, 11}$, particularly with Landau quantization of orbits, we address its role in the statistical thermodynamics of the Group VI Dichalcogenides jointly with that of temperature, as reflected in the entropy of these systems. (As we consider the magnetic field to be constant there is no magnetization energy contribution in Eq(1.2), $\boldsymbol{M} \cdot \delta \mathbf{B} = 0$; $\mathbf{B}$ is the magnetic field strength, $e$ is the electron charge.)  \\
	\indent The use of the retarded Green's function $G^{ret}$ in the determination of statistical thermodynamic functions stems from the fact that its trace in position--time ($\vec{x}, T$) representation produces the ordinary (``classical") partition function $\widehat{Z}(\beta) = Tr \exp(-\beta H)$ with the substitution $T \rightarrow -i\beta$ as$^{12}$
	\begin{equation}
		\widehat{Z}(\beta) = \int d\vec{x}\  Tr (iG_{T > 0}^{ret} (\vec{x}, \vec{x}; T \rightarrow -i\beta))
	\end{equation}
	where $\beta = 1/k_B T'$ is inverse thermal energy, $k_B$ is the Boltzmann constant, $T'$ is Kelvin temperature, and $Tr$ denotes the trace. This is readily verified using the definition of the Green's function in terms of a time translation operator$^{12}$. Actually, it is the logarithm of the \textit{grand} partition function $Z(\beta)$ for Fermions that is required to determine the Helmholtz Free Energy, $F$, and Wilson's book$^{13}$ reports a clever way to obtain it from the ordinary partition function $\widehat{Z}(\beta)$, as follows ($\mu$ is chemical potential and $E_\gamma$ represents the single--fermion energy spectrum.)
	\begin{equation}
		F - \mu N = -k_B T' \ln Z = -k_B T' \sum_{E_\gamma}\ln (1 + e^{-\beta (E_\gamma - \mu)}),
	\end{equation}
	and writing the $E_\gamma$--summand as an inverse Laplace transform
	\begin{equation}
		B(E) \equiv -k_B T' \ln (1 + e^{-\beta(E - \mu)}) = \int_c \frac{ds}{2\sigma i}\ e^{sE} p(s)
	\end{equation}
	with $p(s)$ as the Laplace transform of $B(E)$
	\begin{equation}
		p(s) = \int_0^{\infty}dE\ e^{-sE} B(E),
	\end{equation}
	we have
	\begin{equation}
		F - \mu N = \int_c \frac{ds}{2\pi i}\ p(s) \sum_{E_\gamma} e^{sE_\gamma} = \int_c \frac{ds}{2\pi i}\ p(s) \widehat{Z}(\beta \rightarrow -s).
	\end{equation}
	This expresses the Helmholtz Energy in terms of the ordinary partition function, or alternatively, the Green's function. Wilson further noted that rewriting Eq(1.7) in the form
	\begin{equation}
		F - \mu N= \int_c \frac{ds}{2\pi i}\ \left( \frac{\widehat{Z} (\beta \rightarrow -s)}{s^2} \right) \left( s^2 p(s) \right),
	\end{equation}
	one may employ a useful special case of the convolution theorem for Laplace transforms$^{14}$ to obtain
	\begin{equation}
		F - \mu N = \int_0^\infty dE\ \int_c \frac{ds}{2\pi i}\ e^{Es} \frac{\widehat{Z}(s)}{s^2} \int_c \frac{ds'}{2\pi i}\ e^{Es'} s^{'2} p(s');
	\end{equation}
	and since
	\begin{equation}
		\int_c \frac{ds'}{2\pi i}\  e^{Es'} s^{'2} p(s') = \frac{\partial^2}{\partial E^2} \int_c \ \frac{ds'}{2\pi i}\ e^{Es'} p(s') = \frac{\partial^2 B(E)}{\partial E^2} = \frac{\partial f_0 (E)}{\partial E},
	\end{equation}
	we have the convenience of dealing directly with the temperature dependent Fermi distribution $f_0(E)$ (rather than $B(E)$):
	\begin{equation*}
		F - \mu N = \int_c \frac{ds}{2\pi i}\ \frac{\widehat{Z}(s)}{s^2} \int_0^\infty dE\ e^{Es} \frac{\partial f_0(E)}{\partial E}
	\end{equation*}
	\begin{equation}
		\qquad \qquad \qquad \qquad \qquad = - \frac{\beta}{4} \int_c \frac{ds}{2\pi i}\ \frac{\widehat{Z}(s)}{s^2} \int_0^\infty dE\ e^{Es} \text{sech}^2\left( \frac{[E - \mu] \beta}{2} \right).
	\end{equation}
	
	\chapter{Retarded Green's Function of Group VI Dichalcogenides in a Magnetic Field}
	\addcontentsline{toc}{chapter}{Retarded Green's Function of Group VI Dichalcogenides}
	\indent As indicated above, our approach to the determination of the role of a quantizing magnetic field in the entropy of charge carriers of the Group VI Dichalcogenides in the low energy regime (in which their Hamiltonian is ``Dirac--like" with energy proportional to momentum) is undertaken using the retarded Green's function; In earlier work$^{15}$, the associated Landau--quantized Green's function matrix was derived with full account of its pseudospin--$1/2$ and spin--$1/2$ features in the presence of a high magnetic field; and the pertinent diagonal elements of its retarded Green's function pseudospin matrix $G_{\substack{1 1 \\ 2 2}}^{ret}$ are given in 2D--position ($\vec{R} = \vec{x} - \vec{x}'$) / time ($T = t - t'$) representation as
	\begin{equation*}
	iG_{\substack{1 1 \\ 2 2}}^{ret} (\vec{x}, \vec{x}'; T) = \eta_{+}(T)\exp\left(\frac{ie}{2}[\vec{x} \cdot \vec{B}\times\vec{x}']\right) \frac{eB}{2\pi}e^{-iE_{s_z}T}\exp \left(-\frac{eBR^2}{4}\right) \qquad\qquad\qquad
	\end{equation*}
	\begin{equation}
		\times \sum_{n=0}^{\infty} L_n \left(\frac{eBR^2}{2}\right) \left\lbrace \cos \left( \sqrt{g^2 + \epsilon_{n\pm}^2}T \right)  \mp \frac{ig}{\sqrt{g^2 + \epsilon_{n\pm}^2}} \sin \left(   \sqrt{g^2 + \epsilon_{n\pm}^2}T \right) \right\rbrace .
	\end{equation}
	Here, $\eta_+(T) = 1$ for $T > 0;\  0$ for $T < 0$ is the Heaviside unit step function. A spin index $s_z = \pm 1$ enters into energy shifts as $E_{s_z} = s_z \nu \frac{\lambda}{z}$ with $\nu = \pm 1$ as the valley index; furthermore, $\lambda$ is the spin splitting and $g = \frac{\Delta}{2} - E_{s_z}$ with $\Delta$ as the energy gap without spin splitting. Also, $L_n$ represents the Laguerre polynomials and $\epsilon_{n\pm}^2$ is given by ($1_\nu \equiv sign(\nu) \equiv \pm 1$)
	\begin{equation}
	\epsilon_{n\pm}^2 = (2n + 1\mp1_\nu)\gamma^2 e\mathbf{B},
	\end{equation}
	and $\gamma$ is an effective speed determined by the tight binding hopping parameter and lattice spacing. It is useful to rewrite the trace of Eq(3.1) in the following form (for use below):
	\begin{equation*}
		Tr \left( iG^{ret}_{T > 0}(\vec{x}, \vec{x}'; T) \right) = \exp \left( \frac{ie}{2} [\vec{x} \cdot \vec{B} \times \vec{x}'] \right) \frac{eB}{4\pi} e^{-iE_{s_z}T} \exp\left( \frac{-eBR^2}{4} \right)\qquad\qquad\qquad
	\end{equation*}
	\begin{equation}
		\times \sum_{s_z = \pm 1} \sum_{\nu = \pm 1} \sum_{n = 0}^\infty \sum_\pm \sum_{\pm'} L_n \left( \frac{eBR^2}{2} \right) \left( 1 \mp (\pm'1) \frac{g}{\sqrt{g^2 + \epsilon^2_{n\pm}}} \right) e^{\pm' i \sqrt{g^2 + \epsilon_{n\pm}^2}T}.
	\end{equation}
	This trace encompasses sums over the spin index $s_z = \pm 1$, valley index $\nu = \pm 1$, pseudospin index $\pm$, the signature $\pm'$ of exponentials constituting sine and cosine functions and the Laguerre sum index $n = 0 \rightarrow \infty$. For the problem at hand, $\vec{x} \equiv \vec{x}'$, $\vec{R} \equiv 0$ and $\ln(0) \equiv 1$, whence
	\begin{equation*}
	\int d\vec{x}\ Tr \left( iG^{ret}_{T > 0} (\vec{x}, \vec{x}; T) \right) = \text{(area) } \frac{eB}{4\pi} e^{-iE_{s_z}T}\qquad\qquad\qquad\qquad\qquad\qquad\qquad\qquad\qquad
	\end{equation*}
	\begin{equation}
		\times \sum_{s_z = \pm 1} \sum_{\nu = \pm 1} \sum_{n = 0}^\infty \sum_\pm \sum_{\pm'} \left( 1 \mp (\pm') \frac{g}{\sqrt{g^2 + \epsilon_{n\pm}^2}} \right) e^{\pm' i \sqrt{g^2 + \epsilon_{n\pm}^2}T},
	\end{equation}
	where the (area) factor arises from the 2D $d\vec{x}$--integration. \\
	\indent It is also of interest to describe the \textit{thermodynamic} Green's function matrix, $G(\vec{x}, \vec{x}'; T)$, which is characterized by periodicity in imaginary time (period $\tau = -i\beta = -i/k_B T'$) instead of retardation. It may be written in terms of its "greater" ($G^>$) and "lesser" ($G^<$) constituents as$^{12, 15}$
	\begin{equation}
		G(\vec{x}, \vec{x}'; T) = \eta_{+}(T) G^>(\vec{x}, \vec{x}'; T) + \eta_{+}(-T) G^<(\vec{x}, \vec{x}'; T).
	\end{equation}
	The matrix Green's function constituents $G^>$ and $G^<$ define a corresponding spectral weight matrix $A$ as
	\begin{equation}
		G^>(\vec{x}, \vec{x}'; T) - G^<(\vec{x}, \vec{x}'; T) = -iA(\vec{x}, \vec{x}'; T) = -i\int \frac{d\omega}{2\pi}\ e^{-i\omega T} A(\vec{x}, \vec{x}'; \omega).
	\end{equation}
	Here, $G^>(\vec{x}, \vec{x}'; T)$, $G^<(\vec{x}, \vec{x}'; T)$ and $A(\vec{x}, \vec{x}'; T)$ all satisfy the homogeneous counterpart of the Green's function equation; and all involve the same Peierls phase factor $\exp(\frac{ie}{2}[\vec{x} \cdot \vec{B} \times \vec{x}'])$ which we divide out, defining $A'(\vec{x}, \vec{x}'; T)$ by the relation
	\begin{equation}
		A(\vec{x}, \vec{x}'; T) = \exp \left( \frac{ie}{2} [\vec{x} \cdot \vec{B} \times \vec{x}'] \right) A' (\vec{x}, \vec{x}'; T).
	\end{equation}
	The constituent parts of the \textit{thermodynamic} Green's function matrix may be determined from the spectral weight function matrix in \textit{frequency} representation as
	\begin{equation}
		iG'^{^{\left\{ \substack{> \\ <} \right\}}} (\vec{x}, \vec{x'}; \omega) = \left\{\begin{matrix}
			1&- f_0(\omega)\\
			&- f_0(\omega)
		\end{matrix}\right\} A'(\vec{x}, \vec{x}'; \omega),
	\end{equation}
	where $f_0(\omega)$ is the Fermi--Dirac distribution function. Furthermore $A'(\vec{x}, \vec{x}'; \omega)$ may determined from the structure of the \textit{retarded} Green's function $G^\text{ret}$ in frequency representation using the relation$^{12}$ ($\vec{R} = \vec{x} - \vec{x}'$)
	\begin{equation}
		A'(\vec{x}, \vec{x}'; \omega) = -2 Im G^{'\text{ret}}(\vec{R}; \omega),
	\end{equation}
	leading to the result for its diagonal elements $A'_{\substack{11\\22}}(\vec{x}, \vec{x}'; \omega)$ as$^{15}$
	\begin{equation*}
		A'_{\substack{11\\22}}(\vec{x}, \vec{x}'; \omega) = \frac{eB}{2} \exp \left( \frac{-eBR^2}{4} \right) \left( 1 \pm \frac{g}{\omega - E_{s_z}} \right) \sum_{n = 0}^\infty \sum_{\pm'} L_n \left( \frac{eBR^2}{2} \right)\qquad\qquad\qquad\qquad
	\end{equation*}
	\begin{equation}
		\times \delta \left( \omega - E_{s_z} \pm' \sqrt{g^2 + \epsilon_{n\pm}^2} \right). \qquad\qquad\qquad\qquad\qquad\qquad
	\end{equation}
	From this, the diagonal elements of the thermodynamic Green's function may be determined for the Dichalcogenides in a magnetic field using Eq(2.8). In particular, we obtain $Tr G^<(\vec{x}, \vec{x}'; T)$ as
	\begin{equation*}
		-iTr G^<(\vec{x}, \vec{x}'; T) = \exp \left( \frac{ie}{2} [\vec{x} \cdot \vec{B} \times \vec{x}'] \right) \frac{eB}{4\pi} e^{-\frac{eBR^2}{4}} \sum_{s_z = \pm 1} \sum_{\nu = \pm 1} \sum_{n = 0}^\infty \sum_\pm \sum_{\pm'} \qquad\qquad
	\end{equation*}
	\begin{equation}
		\times f_0\left( E_{s_z} \mp' \sqrt{g^2 + \epsilon_{n\pm}^2} \right) L_n \left( \frac{eBR^2}{2} \right) \left( 1 \pm \frac{(\mp'1)g}{\sqrt{g^2 + \epsilon_{n\pm}^2}} \right) e^{-i\left( E_{s_z} \mp' \sqrt{g^2 + \epsilon_{n\pm}^2} T \right)}.\qquad
	\end{equation}
	The density follows as
	\begin{equation*}
		n = -iTr G'_< (\vec{R} = 0; T = 0) = \frac{eB}{4\pi} \sum_{s_z = \pm 1} \sum_{\nu = \pm 1} \sum_{n = 0}^\infty \sum_\pm \sum_{\pm'} \left( 1 \pm \frac{(\mp'1)g}{\sqrt{g^2 + \epsilon_{n\pm}^2}} \right) \qquad\qquad
	\end{equation*}
	\begin{equation}
		\times f_0 \left( E_{s_z} \mp' \sqrt{g^2 + \epsilon_{n\pm}^2} \right).
	\end{equation}
	
	\chapter{Temperature Dependence of the Helmholtz Free Energy: Degenerate and Nondegenerate Regimes}
	\indent The analysis of temperature dependence devolves upon a careful evaluation of the Helmholtz Free Energy, and we first examine this in the approach to the degenerate regime, $\mu \beta \rightarrow \infty$, by rewriting Eq(1.11) with a change of variable, $z = [E - \mu]\beta / 2$. The resulting E-integral of Eq(1.11) given by
	\begin{equation}
		\int_0^\infty dE_{\dots} \equiv \int_0^\infty dE\ e^{Es} \text{sech}^2 \left( \frac{[E - \mu]\beta}{2} \right) = \frac{2}{\beta} e^{s \mu} \int_{-\mu \beta / 2 \rightarrow - \infty}^\infty dz\ e^{\frac{2s}{\beta}z} \text{sech}^2 z,
	\end{equation}
	and in the degenerate regime only the even part of the exponential integrand contributes, $ e^{2sz/\beta} \rightarrow \text{cosh}(2sz/\beta)$, with the result$^{16}$
	\begin{equation}
		\int_0^\infty dE_{\dots} = \frac{2\pi}{\beta^2} \frac{s e^{s\mu}}{\sin(\frac{\pi s}{2\beta})}.
	\end{equation}
	Therefore, we obtain Eq(1.11) as
	\begin{equation}
		F - \mu N = - \frac{\pi}{2\beta} \int_c \frac{ds}{2\pi i}\ \frac{e^{s\mu} \widehat{Z}(s)}{s \sin(\pi s/2\beta)}.
	\end{equation}
	Employing Eqns(1.3 and 2.4) and noting our successive changes in the argument of $\widehat{Z}$ (summarized as $T \rightarrow -is$ in the Green's function trace; also note that $L_n(0) \equiv 1$ and set $s = \beta s'$), we have
	\begin{equation*}
		F - \mu n = - \frac{eB}{8 \beta} \sum_{s_z = \pm 1} \sum_{\nu = \pm 1} \sum_\pm \sum_{\pm'} \sum_{n = 0}^\infty \left( 1 \mp (\pm') \frac{g}{\sqrt{g^2 + \epsilon^2_{n\pm}}} \right)
	\end{equation*}
	\begin{equation}
		\times \int_c \frac{ds'}{2\pi i}\  \frac{\exp\left[s' \beta \left( \mu - E_{s_z} \pm' \sqrt{g^2 + \epsilon_{n\pm}^2} \right)\right]}{s' \sin(\pi s'/2)}\quad
	\end{equation}
	on a per--unit area basis ($F \rightarrow F/$area and $n = N/$area). This inverse Laplace transform $I = \int_c ds'/2\pi i \cdots$ has contributions from a second order pole at the origin $s' = 0\ (I_0)$, and first order poles at $s' = 2k'$ for all nonvanishing $k'$ integer values $(I_{k'})$ where
	\begin{equation*}
		\frac{1}{\sin(\pi s'/2)} \rightarrow \frac{2}{\pi} \frac{(-1)^{k'}}{s' - 2k'}.
	\end{equation*}
	\indent Considering the second order pole at $s' = 0$, we have
	\begin{equation}
		I_0 \equiv \frac{2}{\pi} \int \frac{ds'}{2\pi i}\ \frac{\exp \left[ s'\beta \left( \mu - E_{s_z} \pm' \sqrt{g^2 + \epsilon_{n\pm}^2} \right) \right]}{s^{'2}},
	\end{equation}
	and closing the $s'$--integration contour in the left--half--plane we obtain
	\begin{equation}
		I_0 = \frac{2}{\pi}\beta \eta_+ \left( \mu - E_{s_z} \pm' \sqrt{g^2 + \epsilon_{n\pm}^2} \right) \left( \mu - E_{s_z} \pm' \sqrt{g^2 + \epsilon_{n\pm}^2} \right);
	\end{equation}
	furthermore, considering the simple poles occurring at $s' = 2k'$ for \textit{positive} integers $k' > 0$, the $s'$--integration contour must be closed in the right--half--plane, so that
	\begin{equation*}
		I_{k' > 0} = -(-1)^{k'} \eta_- \left( \mu - E_{s_z} \pm' \sqrt{g^2 + \epsilon_{n\pm}^2} \right)\qquad\qquad\qquad\quad
	\end{equation*}
	\begin{equation}
		 \times \exp \left[ -2 |k'| \left| \mu - E_{s_z} \pm' \sqrt{g^2 + \epsilon_{n\pm}^2} \right| \beta \right] \bigg/ \pi |k'|,\ 
	\end{equation}
	where $\eta_- (x) = 1$ for $x < 0;\  0$ for $x > 0$ (alternatively, $\eta_- (x) = 1 - \eta_+ (x)$). However, for simple poles occurring at $s' = 2k'$ for \textit{negative} integers $k' < 0$, the contour must again be closed in the left half $s'$--plane, yielding
	\begin{equation*}
		I_{k' < 0} = -(-1)^{k'}\eta_+ \left( \mu - E_{s_z} \pm' \sqrt{g^2 + \epsilon_{n\pm}^2} \right) \qquad\qquad\qquad\qquad\qquad
	\end{equation*}
	\begin{equation}
		\times \exp \left[ -2|k'| \left| \mu - E_{s_z} \pm' \sqrt{g^2 + \epsilon_{n\pm}^2}\right| \beta \right] \bigg/ \pi |k'|. \qquad\quad\ \ 
	\end{equation}
	Collectively, the $s'$--integral sum for all $k'$--integers is given by
	\begin{equation}
		I = \sum_{k' = -\infty}^\infty I_k' = I_0 + \sum_{k' > 0}  I_{k' > 0} + \sum_{k' < 0} I_{k' < 0}
	\end{equation}
	and noting that $\eta_{+}(x) + \eta_{-}(x) = 1$,
	\begin{equation*}
		I = \frac{2}{\pi} \beta \eta_+ \left( \mu - E_{s_z} \pm' \sqrt{g^2 + \epsilon_{n\pm}^2} \right) \left( \mu - E_{s_z} \pm' \sqrt{g^2 + \epsilon_{n\pm}^2} \right)
	\end{equation*}
	\begin{equation}
		- \sum_{\text{all }k' \neq 0} \frac{(-1)^{k'}}{\pi |k'|} \exp \left[ -2|k'| \left| \mu - E_{s_z} \pm' \sqrt{g^2 + \epsilon_{n\pm}^2} \right| \beta \right].\qquad
	\end{equation}
	Thus, in the degenerate regime the resulting temperature--dependent Helmholtz Free Energy is:
	\begin{equation*}
		F - \mu n = -\frac{eB}{4\pi} \sum_{s_z = \pm 1} \sum_{\nu = \pm 1} \sum_\pm \sum_{\pm'} \sum_{n = 0}^\infty \left( 1 \mp (\pm'1) \frac{g}{\sqrt{g^2 + \epsilon^2_{n\pm}}} \right)\qquad\qquad\qquad\qquad\qquad\qquad
	\end{equation*}
	\begin{equation*}
		  \times \left\lbrace \rule{0cm}{.75cm} \eta_+ \rule{-.10cm}{0cm} \left( \mu - E_{s_z} \pm' \sqrt{g^2 + \epsilon_{n\pm}^2} \right) \left[ \mu - E_{s_z} \pm' \sqrt{g^2 + \epsilon_{n\pm}^2} \right] \right. \qquad\qquad\qquad\quad\ \ 
	\end{equation*}
	\begin{equation}
		\left. -\frac{1}{2\beta} \sum_{\text{all }k' \neq 0} \frac{(-1)^{k'}}{|k'|} \exp \left[ -2|k'| \left| \mu - E_{s_z} \pm' \sqrt{g^2 + \epsilon_{n\pm}^2} \right| \beta \right] \right\rbrace.\qquad\qquad
	\end{equation}
	(Since the $k'$--summand of Eq(3.11) is insensitive to the sign of $k'$, it could also be written as $\sum_{\text{all } k' \neq 0} = 2\sum_{k' > 0}$.) This result for $F - \mu n$ in the degenerate regime embodies the zero--temperature limit (first term in the curly bracket) and exponentially small temperature corrections (last term) in the approach to zero temperature. This temperature dependent last term involving the series $\sum_{\text{all } k' \neq 0} = 2\sum_{k' > 0}$ may be summed in closed form since it is related to a simple geometric series as follows:
	\begin{equation*}
		\frac{\partial}{\partial \beta} \rule{-.15cm}{0cm}\left(\rule{-.10cm}{0cm} \sum_{\text{all }k' \neq 0}\rule{-.20cm}{0cm} \frac{(-1)^{k'}}{|k'|} \exp \rule{-.15cm}{0cm}\left[ -2|k'| \left| \mu - E_{s_z} \pm' \sqrt{g^2 + \epsilon_{n\pm}^2} \right|\rule{-.10cm}{0cm} \beta \right] \rule{-.20cm}{0cm} \right)\rule{-.20cm}{0cm} =\rule{-.10cm}{0cm} -2 \left| \mu - E_{s_z} \pm' \sqrt{g^2 + \epsilon_{n\pm}^2} \right| \Gamma
	\end{equation*}
	where $\Gamma$ is proportional to the geometric series (less the unity term)
	\begin{equation*}
		\Gamma \equiv 2\sum_{k' > 0} (-1)^{k'} \exp \left[ -2k' \left| \mu - E_{s_z} \pm' \sqrt{g^2 + \epsilon_{n\pm}^2} \right| \beta \right],
	\end{equation*}
	which may be summed as
	\begin{equation}
		\Gamma = -2 \left( \exp \left[ 2 \left| \mu - E_{s_z} \pm' \sqrt{g^2 + \epsilon_{n\pm}^2} \right| \beta \right] + 1 \right)^{-1}.
	\end{equation}
	Integration of Eq(3.12) with respect to $\beta$ from $\beta$ to $\infty$ yields$^{17}$
	\begin{equation*}
		\sum_{\text{all }k' \neq 0} \frac{(-1)^{k'}}{|k'|} \exp \left[ -2|k'| \left| \mu - E_{s_z} \pm' \sqrt{g^2 + \epsilon_{n\pm}^2} \right| \beta \right]
	\end{equation*}
	\begin{equation}
		 = -2\ln \left( 1 + \exp \left[ -2 \left| \mu - E_{s_z} \pm' \sqrt{g^2 + \epsilon_{n\pm}} \right| \beta \right] \right).
	\end{equation}
	Consequently, we obtain the result\\ \ \\ \ \\ \ \\
	\underline{\textbf{DEGENERATE REGIME}}
	\begin{equation*}
	F - \mu n = -\frac{eB}{4\pi} \sum_{s_z = \pm 1} \sum_{\nu = \pm 1} \sum_\pm \sum_{\pm'} \sum_{n = 0}^\infty \left( 1 \mp (\pm'1) \frac{g}{\sqrt{g^2 + \epsilon^2_{n\pm}}} \right)\qquad\qquad\qquad\qquad\qquad\qquad
	\end{equation*}
	\begin{equation*}
	\quad\;  \times \left\lbrace \rule{0cm}{.75cm}\eta_+\rule{-.10cm}{0cm} \left( \mu - E_{s_z} \pm' \sqrt{g^2 + \epsilon_{n\pm}^2} \right) \left[ \mu - E_{s_z} \pm' \sqrt{g^2 + \epsilon_{n\pm}^2} \right] \right. \qquad\qquad\qquad\quad\ \ 
	\end{equation*}
	\begin{equation}
	\left. + \frac{1}{\beta} \ln \rule{-.10cm}{0cm}\left( 1 + \exp\left[ -2 \left| \mu - E_{s_z} \pm' \sqrt{g^2 + \epsilon_{n\pm}^2} \right| \beta \right] \right) \rule{0cm}{.75cm}\right\}. \qquad\qquad\qquad\ \ 
	\end{equation}
	\indent The nondegenerate regime is readily obtained as the leading term in the fugacity expansion of $F - \mu n$ in powers of $\exp[(\mu - E)\beta]$, so the $E$--integral of Eq(3.1) takes the form
	\begin{equation}
		\int_0^{\infty} dE_{\cdots} = 4e^{\mu \beta} \int_0^{\infty} dE\ e^{E(s - \beta)} = -\frac{4e^{\mu \beta}}{s - \beta}
	\end{equation}
	and consequently
	\begin{equation*}
		F - \mu n = \frac{\beta eB}{4\pi} e^{\mu \beta} \sum_{s_z = \pm 1} \sum_{\nu = \pm 1} \sum_{n = 0}^{\infty} \sum_{\pm} \sum_{\pm'} \left( 1 \mp (\pm') \frac{g}{\sqrt{g^2 + \epsilon^2_{n\pm}}} \right)
	\end{equation*}
	\begin{equation}
		\times \int_c \frac{ds}{2\pi i}\ \frac{1}{s^2(s - \beta)} e^{s\left( -E_{s_z} \pm' \sqrt{g^2 + \epsilon^2_{n\pm}} \right)}.
	\end{equation}
	Noting the second order pole at $s = 0$ and the first order pole at $s = \beta > 0$, we obtain the nondegenerate result for the inverse Laplace transform as
	\begin{equation*}
		\int_c \frac{ds}{2\pi i} \cdots = - \frac{1}{\beta} \eta_+ \left( -E_{s_z} \pm' \sqrt{g^2 + \epsilon^2_{n\pm}} \right) \left[ -E_{s_z} \pm' \sqrt{g^2 + \epsilon^2_{n\pm}} \right]
	\end{equation*}
	\begin{equation}
		- \frac{1}{\beta^2} \eta_- \left( -E_{s_z} \pm' \sqrt{g^2 + \epsilon^2_{n\pm}} \right) e^{- \beta \left| -E_{s_z} \pm' \sqrt{g^2 + \epsilon^2_{n\pm}} \right|},
	\end{equation}
	Finally, the nondegenerate result for $F - \mu n$ (per unit area) is given by\\ \ \\
	\underline{\textbf{NONDEGENERATE REGIME}}
	\begin{equation*}
		F - \mu n = - \frac{eB}{4 \pi} e^{\mu \beta} \sum_{s_z = \pm 1} \sum_{\nu = \pm 1} \sum_{n = 0}^{\infty} \sum_{\pm} \sum_{\pm'} \left( 1 \mp (\pm'1) \frac{g}{\sqrt{g^2 + \epsilon^2_{n\pm}}} \right)\qquad\qquad\qquad\qquad\qquad
	\end{equation*}
	\begin{equation*}
		\times \left\lbrace \eta_+ \rule{-.1cm}{0cm} \left( -E_{s_z} \pm' \sqrt{g^2 + \epsilon^2_{n\pm}} \right) \left[ -E_{s_z} \pm' \sqrt{g^2 + \epsilon^2_{n\pm}} \right] \right.\qquad\qquad\qquad\qquad\ 
	\end{equation*}
	\begin{equation}
			\left. + \frac{1}{\beta} \eta_- \rule{-.1cm}{0cm} \left( -E_{s_z} \pm' \sqrt{g^2 + \epsilon^2_{n\pm}} \right) e^{- \beta \left| -E_{s_z} \pm' \sqrt{g^2 + \epsilon^2_{n\pm}} \right|} \right\rbrace,\qquad\qquad\qquad
	\end{equation}
	where $\eta_{-} (x) = 1 - \eta_{+} (x)$, as defined above.
	
	\chapter{Entropy and Magnetic Moment of the Group VI Dichalcogenides: Temperature Dependence in the Degenerate and Nondegenerate Statistical Regimes}
	\indent The determination of the entropy $S$ (per unit area) is made using Eq(1.2) in the form
	\begin{equation}
		S =- \frac{\partial\beta}{\partial T'} \frac{\partial(F - \mu n)}{\partial\beta} = k_B \beta^2 \frac{\partial(F - \mu n)}{\partial\beta}.
	\end{equation}
	Of course, the first statement to make is that in the zero temperature limit, $\beta \rightarrow \infty$, the entropy is expected to vanish: This is readily verified from Eq(3.3), whose low temperature limit is given by
	\begin{equation*}
		[F - \mu n]_{\beta >> 1} = -\int_c \frac{ds}{2\pi i}\ \frac{e^{s\mu}\widehat{Z}(s)}{s^2} + 0(\frac{1}{\beta^2})\text{ for $\beta >> 1$,}
	\end{equation*}
	whence
	\begin{equation}
		 S_{T' \rightarrow 0} = -\frac{\partial}{\partial T'}[F - \mu n] \equiv 0.
	\end{equation}
	An alternative, fully general, proof based on Eqns(1.3--1.5) follows:
	\begin{equation*}
	S = \frac{\partial\beta}{\partial T'} \frac{\partial}{\partial\beta} \left( \beta^{-1} \sum_{E_\gamma} \ln \left( 1 + e^{-\beta (E_\gamma - \mu)} \right) \right)\qquad\qquad\qquad\qquad\qquad\qquad\qquad\qquad\qquad\qquad
	\end{equation*}
	\begin{equation*}
	= k_B \beta^2 \sum_{E_\gamma} \left\lbrace \beta^{-2} \ln \left( 1 + e^{-\beta[E_\gamma - \mu]} \right) + \beta^{-1} \frac{e^{-\beta [E_\gamma - \mu]}[E_\gamma - \mu]}{1 + e^{-\beta [E_\gamma - \mu]}} \right\rbrace \qquad\qquad\qquad\quad\ \ \ 
	\end{equation*}
	\begin{equation}
	= k_B \sum_{E_\gamma} \ln \left( 1 + e^{-\beta [E_\gamma - \mu]} \right) + k_B \beta \sum_{E_\gamma} [E_\gamma - \mu] f_0 (E_\gamma - \mu). \qquad\qquad\qquad\qquad\quad
	\end{equation}
	and in the limit of zero temperature ($\beta \rightarrow \infty$),
	\begin{equation*}
	S = k_B \sum_{E_\gamma} \left(\eta_{+} (E_\gamma - \mu) \ln(1) + \eta_{+} (\mu - E_\gamma) \ln\left( e^{-\beta[E_\gamma - \mu]} \right)\right)
	\end{equation*}
	\begin{equation}
	+ k_B \beta \sum_{E_\gamma} [E_\gamma - \mu] \left[ \eta_{+} (\mu - E_\gamma) + \eta_{+}(E_\gamma - \mu)(0) \right],\qquad
	\end{equation}
	whence
	\begin{equation}
	S = \sum_{E_\gamma} (k_B \beta - k_B \beta) \eta_{+}(\mu - E_\gamma) [E_\gamma - \mu] \equiv 0. \qquad\qquad\qquad\quad\ 
	\end{equation}
	It is worthwhile to review this matter because we are dealing with a system model that has an unbounded negative energy component (as well as a positive one) in its spectrum. \\
	\indent Furthermore, we examine the entropy $S_\text{Deg}$ in the approach to zero temperature, ie: the degenerate regime: Employing Eq(3.14) and Eq(4.1) we find
	\begin{equation*}
		S_\text{Deg} = k_B \frac{eB}{4\pi} \sum_{s_z = \pm 1} \sum_{\nu = \pm 1} \sum_{n = 0}^\infty \sum_{\pm} \sum_{\pm'} \left( 1 \mp (\pm') \frac{g}{\sqrt{g^2 + \epsilon^2_{n\pm}}} \right)\qquad\qquad\qquad\qquad\qquad\qquad
	\end{equation*}
	\begin{equation*}
		\times \left\lbrace \ln \left( 1 + \exp \left[ -2 \left| \mu - E_{s_z} \pm' \sqrt{g^2 + \epsilon_{n\pm}^2} \right| \beta \right] \right) \right. \qquad\qquad\qquad\qquad\qquad\qquad\qquad\qquad
	\end{equation*}
	\begin{equation}
		+ \left. 2 \beta \left| \mu - E_{s_z} \pm' \sqrt{g^2 + \epsilon_{n\pm}^2} \right| \left[ 1 + \exp\left( 2 \left| \mu - E_{s_z} \pm' \sqrt{g^2 + \epsilon_{n\pm}^2} \right| \beta \right) \right]^{-1} \right\rbrace. \qquad\qquad\qquad
	\end{equation}
	Clearly, $S_{\text{Deg}}$ vanishes exponentially as $T' \rightarrow 0$, as expected. \\
	\indent In regard to the nondegenerate regime, the entropy may be determined using Eq(3.15), with the result
	\begin{equation*}
		S_\text{Nondeg} = -k_B \beta^2 \frac{eB}{4\pi} e^{\mu\beta} \sum_{s_z = \pm 1} \sum_{\nu = \pm 1} \sum_{n = 0}^\infty \sum_{\pm} \sum_{\pm'}\left( 1 \mp (\pm') \frac{g}{\sqrt{g^2 + \epsilon^2_{n\pm}}} \right)\qquad\qquad\qquad\qquad
	\end{equation*}
	\begin{equation*}
		\times \left\lbrace \mu \eta_{+}\rule{-.10cm}{0cm} \left( -E_{s_z} \pm' \sqrt{g^2 + \epsilon^2_{n\pm}} \right) \left[ -E_{s_z} \pm' \sqrt{g^2 + \epsilon^2_{n\pm}} \right] \right.\qquad\qquad\qquad\qquad\qquad\qquad\qquad
	\end{equation*}
	\begin{equation}
		+\rule{-.1cm}{0cm} \left. \eta_{-}\rule{-.1cm}{0cm} \left( -E_{s_z} \pm' \sqrt{g^2 + \epsilon^2_{n\pm}} \right) e^{-\beta \left| -E_{s_z} \pm' \sqrt{g^2 + \epsilon^2_{n\pm}} \right|} \left( \frac{\mu}{\beta} - \frac{1}{\beta^2} - \frac{1}{\beta} \left| E_{s_z} \pm' \sqrt{g^2 + \epsilon^2_{n\pm}} \right| \right)\rule{-.15cm}{0cm} \right\rbrace.
	\end{equation}
	\indent The results exhibited in Eqns.(3.11) and (3.15) (for the Helmholtz Free Energy and in the degenerate and nondegenerate regimes, respectively) and in Eqns(4.6) and (4.7) (for the entropy in the degenerate and nondegenerate regimes, respectively) clearly exhibit the effects of the quantizing magnetic field jointly with those of finite temperature. To elaborate further on the role of the magnetic field we also evaluate the magnetic moment, $M$, which may be obtained from the free energy as (per unit area)
	\begin{equation}
		M = - \frac{\partial F}{\partial B} = -\frac{\partial(F-\mu n)}{\partial B} - \mu \frac{\partial n}{\partial B} \equiv \Delta M - \frac{\partial n}{\partial B},
	\end{equation}
	where we have defined $\Delta M \equiv -\partial(F - \mu n)/\partial B$. Employing Eq(3.11) for the degenerate regime, we obtain
	\begin{equation*}
		\Delta M_{\text{Deg}} = \frac{e}{4\pi}\sum_{s_z=\pm 1}\sum_{\nu=\pm 1}\sum_{n=0}^\infty\sum_\pm\sum_{\pm'}\left\{\rule{0cm}{.8cm}\left( 1\mp \frac{(\pm'1)g}{\sqrt{g^2+\epsilon_{n\pm}^2}} \pm \frac{(\pm'1)g\epsilon_{n\pm}^2}{2(g^2 + \epsilon_{n\pm}^2)^{3/2}} \right) \right.\qquad\qquad\qquad\qquad
	\end{equation*}
	\begin{equation*}
		\times \left[ \rule{0cm}{.75cm} \eta_{+}\left(\mu - E_{s_z}\pm' \sqrt{g^2 + \epsilon_{n\pm}^2}\right)\left( \mu - E_{s_z}\pm' \sqrt{g^2 + \epsilon_{n\pm}^2} \right) \right. \qquad\qquad\qquad\qquad\qquad\qquad\qquad
	\end{equation*}
	\begin{equation*}
		+ \left. \frac{1}{\beta} \ln \left( 1 + e^{-2\beta \left| \mu - E_{s_z} \pm' \sqrt{g^2 + \epsilon_{n\pm}^2}  \right|} \right) \rule{0cm}{.75cm} \right] \qquad\qquad\qquad\qquad\qquad\quad
	\end{equation*}
	\begin{equation*}
	 	+ \left( 1 \mp \frac{(\pm'1)g}{\sqrt{g^2 + \epsilon_{n\pm}^2}} \right) \left[\rule{0cm}{.9cm} \eta_{+}\rule{-.1cm}{0cm} \left( \mu - E_{s_z} \pm' \sqrt{g^2 + \epsilon_{n\pm}^2} \right) \frac{(\pm'1) \epsilon_{n\pm}^2}{2\sqrt{g^2 + \epsilon_{n\pm}^2}} \right. \qquad\qquad\qquad\qquad\qquad
	\end{equation*}
	\begin{equation}
		 \left. \left. - \frac{\epsilon_{n\pm}^2 \bigg/ \sqrt{g^2 + \epsilon_{n\pm}^2}}{1 + \exp \left[ 2 \left| \mu - E_{s_z} \pm' \sqrt{g^2 + \epsilon_{n\pm}^2} \right| \beta \right]} \right] \rule{0cm}{.8cm}\right\}. \qquad\qquad\qquad\qquad
	\end{equation}
	\indent Furthermore, in the nondegenerate regime Eq(3.15) leads to
	\begin{equation*}
		\Delta M_\text{Nondeg} = \frac{e}{4\pi}e^{\mu\beta} \sum_{s_z=\pm 1}\sum_{\nu=\pm 1}\sum_{n=0}^\infty \sum_\pm \sum_{\pm'} \left[\left( 1\mp\frac{(\pm'1)g}{\sqrt{g^2+\epsilon_{n\pm}^2}} \pm \frac{(\pm'1)g\epsilon_{n\pm}^2}{2(g^2+\epsilon_{n\pm}^2)^{3/2}} \right)\right.\qquad
	\end{equation*}
	\begin{equation*}
		\times \left\{\rule{0cm}{.75cm}\right. \eta_{+}\left( -E_{s_z}\pm'\sqrt{g^2+\epsilon_{n\pm}^2} \right) \left[ -E_{s_z}\pm'\sqrt{g^2+\epsilon_{n\pm}^2} \right]\qquad\qquad\qquad\qquad\qquad\qquad\qquad
	\end{equation*}
	\begin{equation*}
		+ \frac{1}{\beta}\eta_{-}\left( -E_{s_z}\pm'\sqrt{g^2+\epsilon_{n\pm}^2} \right) e^{-\beta\left| -E_{s_z}\pm'\sqrt{g^2+\epsilon_{n\pm}^2} \right|} \left\}\rule{0cm}{.75cm}\right.\qquad\qquad\quad
	\end{equation*}
	\begin{equation*}
		+ \left( 1\mp \frac{(\pm'1)g}{\sqrt{g^2+\epsilon_{n\pm}^2}} \right) \left( \frac{\epsilon_{n\pm}^2}{2\sqrt{g^2+\epsilon_{n\pm}^2}} \right)\qquad\qquad\qquad\qquad\qquad\qquad\qquad\qquad\qquad\qquad
	\end{equation*}
	\begin{equation*}
		\left.\times \left\{\rule{-.2cm}{.75cm}\right. (\pm'1)\eta_{+}\rule{-.1cm}{0cm}\left( -E_{s_z}\pm'\sqrt{g^2+\epsilon_{n\pm}^2} \right)\rule{-.175cm}{0cm}-\eta_{-}\rule{-.1cm}{0cm}\left( -E_{s_z}\pm'\sqrt{g^2+\epsilon_{n\pm}^2} \right) e^{-\beta\left| -E_{s_z}\pm'\sqrt{g^2+\epsilon_{n\pm}^2} \right|} \right.
	\end{equation*}
	\begin{equation}
		 \mp' \beta^{-1} \delta \left( -E_{s_z} \pm' \sqrt{g^2 + \epsilon_{n\pm}^2} \right) \left.\left.  \rule{0cm}{.8cm} \right\} \right]. \qquad\qquad\qquad\qquad\quad\ \ \ 
	\end{equation}
	The final term of $\Delta M_\text{Nondeg}$, namely $\mp' \beta^{-1} \delta \left( -E_{s_z} \pm' \sqrt{g^2 + \epsilon_{n\pm}^2} \right)$, indicates a huge magnetic response as the spin--split/displaced ``relativistic" Landau levels are successively driven by the magnetic field $B$ $\left( \epsilon_{n\pm}^2 = [2n + 1\mp 1_\nu] \gamma^2 eB \right)$ to cause a vanishing of the energy argument $E_{s_z} - \sqrt{g^2 + \epsilon_{n\pm}^2} = 0$. Of course, such a huge response will be moderated by scattering and other interactions. \\
	\indent The last term of $M(-\partial n/ \partial B)$ may be evaluated by differentiating Eq(2.12), with the result
	\begin{equation*}
		- \frac{\partial n}{\partial B} = -\frac{e}{4\pi} \sum_{s_z = \pm 1} \sum_{\nu = \pm 1} \sum_{n= 0}^\infty \sum_\pm \sum_{\pm'} \left\{ \left( 1 \pm \frac{(\mp'1)g}{\sqrt{g^2 + \epsilon_{n\pm}^2}} \right) \left[\rule{0cm}{.8cm} f_0 \left( E_{s_z} \mp' \sqrt{g^2 + \epsilon_{n\pm}^2} \right) \right. \right. \qquad
	\end{equation*}
	\begin{equation*}
		\left. - \frac{\beta}{4} \text{sech}^2 \left( \frac{\left[ E_{s_z} \mp' \sqrt{g^2 + \epsilon_{n\pm}^2} - \mu \right]\beta}{2} \right) \cdot \left( \frac{(\mp'1)\epsilon_{n\pm}^2}{2\sqrt{g^2 + \epsilon_{n\pm}^2}} \right) \right]\qquad\qquad
	\end{equation*}
	\begin{equation}
		\left. \pm \frac{(\pm'1)g\epsilon_{n\pm}^2}{2(g^2 + \epsilon_{n\pm}^2)^{3/2}} f_0 \left( E_{s_z} \mp' \sqrt{g^2 + \epsilon_{n\pm}^2} \right) \rule{0cm}{.75cm}\right\rbrace. \qquad\qquad\qquad\qquad\quad
	\end{equation}
	Taken jointly with Eqns(4.9) and (4.10), Eq(4.11) completes the determination of the magnetization, $M = \Delta M - \partial n/\partial B$. Needless to say, the Fermi--Dirac distribution and its derivative at arbitrary temperature ($\partial f_0(E)/\partial E = - (\beta /4) \text{sech}^2([E - \mu]\beta / 2)$) in Eq(4.11) may be taken in the degenerate and nondegenerate (Maxwellian) regimes to explicitly exhibit the behavior of $M$ in those regimes.
	
	\newpage
	\addcontentsline{toc}{chapter}{References}
	\chapter*{References}
	\begin{enumerate}
		\item M. Katsnelson, ``Graphene: Carbon in Two Dimensions," Cambridge University Press (2012).
		\item H. Aoki and M. S. Dresselhaus, ``Physics of Graphene," Springer (2013).
		\item E. L. Wolf, ``Graphene: A New Paradigm in Condensed Matter and Device Physics," (2013).
		\item T. O. Wehling, A. M. Black--Shaffer and A. V. Balatsky, ``Dirac Materials," arXiv: 1405.5774 v1 [cond-mat.mtrl--sci] (22 May 2014).
		\item J. Wang, S. Deng, Zhongfan Liu and Zhirong Liu, ``The Rare 2D Materials with Dirac Cones," National Science Review \textbf{2}:22--39 (2015).
		\item N. J. M. Horing, ``Aspects of the Theory of Graphene," Transactions Royal Society \textbf{A 368}, 5525--56 (2010).
		\item M. J. S. Spencer, ``Silicene," Springer (2016.)
		\item S. Q. Shen, ``Topological Insulators," Springer (2012).
		\item G. K. Ahluwalia, Editor: ``Applications of Chalcogenides: S, So, Te," Springer (2017).
		\item D. Lei, ``Matter in Strong Magnetic Fields," \textit{Rev. Mod. Phys.} \textbf{73}, 629 (2001).
		\item National Research Council, ``High Magnetic Field Science and Its Applications in the United States: Current Status and Future Directions," Washington DC, National Academic Press (2013).
		\item N. J. M. Horing, ``Quantum Statistical Field Theory," Oxford University Press (2017).
		\item A. H. Wilson, ``Theory of Metals," Cambridge University Press, 2nd Edition, Section 6.6 (1965).
		\item A. Erdelyi, Editor, Tables of Integral Transforms 1, p. 131 \# 21 (1954).
		\item N. J. M. Horing, ``Landau Quantized Dynamics and Spectra for Group VI Dichalcogenides, Including a Model Quantum Wire," AIP Advances 7, 065316 (2017).
		\item Reference 14, p. 30 \#2.
		\item H. Dwight, ``Tables of Integrals and Other Mathematical Data," Macmillan Co., p. 127 \#569 (1947).
	\end{enumerate}
	
\end{document}